\documentclass[]{jfm_preprint}
\usepackage[utf8]{inputenc}
\usepackage[T1]{fontenc}
\usepackage{graphicx}
\usepackage{newtxtext}
\usepackage{newtxmath}
\usepackage{bm}
\usepackage{natbib}
\usepackage{hyperref}
\hypersetup{
    colorlinks = true,
    urlcolor = blue,
    citecolor = black,
    linkcolor = black,
}

\newcommand{\RomanNumeralCaps}[1]

 \usepackage{xcolor}

\newcommand{\rev}[1]{\textcolor{black}{#1}}

\title{Vortex line entanglement in active Beltrami flows}

\author{
    Nicolas Romeo\aff{1,2}, 
    Jonasz S\l{}omka\aff{3}, 
    J\"orn Dunkel\aff{1}
    \and Keaton J. Burns\aff{1,4}\corresp{\email{kjburns@mit.edu}}
 }

\affiliation{
    \aff{1}Department of Mathematics, Massachusetts Institute of Technology, USA
    \aff{2}Department of Physics, Massachusetts Institute of Technology, USA
    \aff{3}Institute of Environmental Engineering, ETH Zürich, CH
    \aff{4}Center for Computational Astrophysics, Flatiron Institute, USA
}

\begin{document}
\maketitle

\begin{abstract}
Over the last decade, substantial progress has been made in understanding the topology of quasi-2D non-equilibrium fluid flows driven by ATP-powered microtubules and microorganisms. 
By contrast, the topology of 3D active fluid flows still poses interesting open questions.
Here, we study the topology of a spherically confined active flow using 3D direct numerical simulations of generalized Navier-Stokes (GNS) equations at the scale of typical microfluidic experiments. 
Consistent with earlier results for unbounded periodic domains, our simulations confirm the formation of Beltrami-like bulk flows with spontaneously broken chiral symmetry in this model. 
Furthermore, by leveraging fast methods to compute linking numbers, we explicitly connect this chiral symmetry breaking to the entanglement statistics of vortex lines.
We observe that the mean of linking number distribution converges to the global helicity, consistent with the asymptotic result by Arnold.
Additionally, we characterize the rate of convergence of this measure with respect to the number and length of observed vortex lines, and examine higher moments of the distribution.
We find that the full distribution is well described by a k-Gamma distribution, in agreement with an entropic argument. 
\rev{Beyond active suspensions, the tools for the topological characterization of 3D vector fields developed here are applicable to any solenoidal field whose curl is tangent to or cancels at the boundaries in a simply connected domain.}
\end{abstract}


\section{Introduction}
\label{sec:intro}

Active turbulence, similar to its passive classical counterpart, is characterized by the emergence of highly complex bulk flow dynamics~\citep{alert_active_2022, matsuzawa_creation_2023,bentkamp_statistical_2022}.
Active fluids based on motile bacteria \citep{sokolov_concentration_2007,wensink_meso-scale_2012,dunkel_fluid_2013}, molecular motors \citep{sanchez_spontaneous_2012}, and self-propelled colloids \citep{bricard_emergence_2013} can display a rich set of topological structures, from spontaneously forming and annihilating point-defects in 2D films \citep{doostmohammadi_stabilization_2016} to entangled vortex lines in 3D bulk flows \citep{copar_topology_2019}. 
Building on classic work on the statistical mechanics of point defects \citep{onsager_statistical_1949,kosterlitz_ordering_1973}, the dynamics and statistics of topological defects have been extensively studied in (quasi) two-dimensional (2D) active fluids \citep{thampi_vorticity_2014,giomi_geometry_2015,james_turbulence_2018, chardac_topology-driven_2021}. 
By contrast, the diverse and complex singular structures realized by active flows in three-dimensional (3D) space \citep{binysh_three-dimensional_2020} were until recently inaccessible to experimental and numerical studies. 
With modern experimental imaging techniques \citep{duclos_topological_2020} and simulation methods, it is now possible to probe 3D topological structures and their statistics \citep{kralj_defect_2023}.

Topological approaches have helped progress theoretical fluid mechanics since the observation by \citet{moffatt_degree_1969} that the tangling of vortex lines is related to the total helicity of the associated ideal incompressible flow.
The total helicity $H$ is an inviscid invariant of incompressible flow defined by the integral \citep{moffatt_degree_1969}
\begin{equation}
    H(t) = \int_V \mathrm{d}^3\bm{x} \,\bm{u}(\bm{x},t)\bm{\cdot}\boldsymbol{\omega}(\bm{x},t),
\end{equation}
where $\bm{u}$ is the flow velocity and $\boldsymbol{\omega} = \nabla \times \bm{u}$ is its associated vorticity. 
Moffatt showed that for a system of $N$ closed and isolated vortex lines in simply connected domains, the total helicity can be expressed as
\begin{equation}
    H = \sum_{i,j=1}^N \kappa_i \kappa_j \mathcal{L}_{ij} \label{eq:helicity_moffatt}
\end{equation}
where $\kappa_i$ is the circulation around the $i$-th vortex tube, and $\mathcal{L}_{ij}$ is the linking number between the centrelines of the $i$- and $j$-th vortex tubes, a topological measure counting the signed integer number of times the $j$-th vortex tube wraps around the $i$-th. 
This connection between helicity and flow topology has been measured in hydrodynamic experiments \citep{kleckner_creation_2013,scheeler_complete_2017} and applies generally to solenoidal fields with discrete localization of their curl, including magnetic fields \citep{berger_topological_1984} or quantum flows \citep{hanninen_vortex_2014, zuccher_relaxation_2017}.

Intuitively, $H$ measures the winding of vortex lines around each other, and non-zero helicities indicate chiral flows in which vortex lines curl around each other in a preferred orientation. 
While this connection between helicity and topology is far-reaching, in most flows of interest, the topology of the vorticity field is much more complex than a set of potentially interlinked discrete vortex tubes. 
As turbulence emerges, experimental resolution of vortex lines becomes more challenging in the absence of localized vortex tubes \citep{matsuzawa_creation_2023}, and equation \eqref{eq:helicity_moffatt} does not directly hold: for general vorticity fields, vortex lines are not necessarily closed, and an asymptotic formulation of equation~\eqref{eq:helicity_moffatt} due to \cite{arnold_asymptotic_1974} provides an extension of Moffatt's result. 
Although theoretically useful, Arnold's formulation has been relatively underutilized in practice due to the high computational cost involved in the computation of the relevant topological quantities. 
It is thus desirable to establish practical means to characterize the topological structure of diffuse vorticity fields, \rev{independently of the dynamics through which these arise}.

Here, building on these core ideas, we study the statistics of vortex line entanglement in a turbulent flow governed by the incompressible generalized Navier Stokes (GNS) equations \citep{slomka_spontaneous_2017,slomka_nature_2018,supekar_linearly_2020}
\begin{subequations}
\begin{align}
    \partial_t \bm{u} + (\bm{u}\bm{\cdot} \nabla) \bm{u} &= -\nabla p + \Gamma_0 \nabla^2 \bm{u}-\Gamma_2 \nabla^4 \bm{u} + \Gamma_4 \nabla^6 \bm{u}\\
    \nabla \bm{\cdot} \bm{u} & = 0 
\end{align} \label{eq:GNS_all}%
\end{subequations}
which model flows with advected active constituents driving a generic linear instability \citep{rothman_negative-viscosity_1989,beresnev_model_1993,tribelsky_new_1996,linkmann_condensate_2020}. 
\rev{While advection is nominally negligible in the low-Reynolds number regime typical of microfluidic experiments, the presence of active stresses in non-dilute microswimmer suspensions can significantly alter the viscous balance, effectively cancelling the fluid viscosity~\citep{lopez_turning_2015} and creating an unstable band of modes.
These modes can saturate nonlinearly via the advective term, and the resulting system can therefore have a large effective Reynolds number and exhibit turbulent-like behaviour \citep{dunkel_minimal_2013, linkmann_condensate_2020,koch_role_2021}.}

\rev{The GNS equations model this behaviour by inducing a band-limited linear instability via the $\Gamma$ terms which saturates in a finite-amplitude, statistically stationary flow.}
For $\Gamma_0, \Gamma_4 >0$ and $\Gamma_2 < 0$ the parameters $(\Gamma_0, \Gamma_2, \Gamma_4)$ together define a characteristic energy injection lengthscale $\Lambda$ and bandwidth $\kappa$, along with a characteristic timescale $\tau$:
\begin{align}
    \Lambda = \pi \sqrt{-\frac{2\Gamma_4}{\Gamma_2}}, 
    \qquad
    \kappa = \left(-\frac{\Gamma_2}{\Gamma_4} - 2\sqrt{\frac{\Gamma_0}{\Gamma_4}}\right)^{1/2},
    \qquad
    \tau = \frac{2\Gamma_4}{\Gamma_2} \left(\Gamma_0 - \frac{\Gamma_2^2}{4\Gamma_4} \right)^{-1}.
\end{align}
Physically, $\Lambda$ prescribes the characteristic vortex size, while $\kappa$ determines how many modes with wavelengths around $\Lambda$ are excited by the linear instability.
The phenomenology of typical microfluidic experiments with bacterial suspensions \citep{dunkel_fluid_2013,wioland_ferromagnetic_2016,copar_topology_2019,peng_imaging_2021} can be recovered by setting $\Lambda \sim 75$ $\mu$m, $\kappa \sim 30$ mm$^{-1}$, and with a characteristic speed $U = 2\pi \Lambda/\tau \sim 10$ $\mu$m/s \citep{slomka_spontaneous_2017}.

While previous analytical and numerical studies of Eq.~\eqref{eq:GNS_all} and other active fluids in 3D have focused on unbounded domains~\citep{slomka_spontaneous_2017,urzay_multi-scale_2017, slomka_nature_2018}, analysing topological structures in a typical experimental setup requires confined simulations. 
Moreover, Moffatt's and Arnold's theorems, like many theoretical tools relying on topological properties of the ambient space, are technically applicable only in simply connected domains. 
We hence choose a 3D ball of radius $R$ as our simulation domain, which is qualitatively similar to experimentally realisable microfluid cavities \citep{wioland_confinement_2013}.

In this work, we leverage spectral direct numerical simulations \citep{burns_dedalus_2020} and recent methods to efficiently compute linking numbers \citep{qu_fast_2021} to explore the topological structure of a numerical realization of active turbulence in confinement (figure~\ref{fig:fig_3d}). 
\rev{However, the methodology developed here is applicable to any incompressible flow, including passive and active fluids.} 
Our active model spontaneously breaks spatial parity and, as previously shown in periodic domains, produces a quasi-Beltrami flow in the bulk. 
These findings are presented in \S~\ref{sec:chiral}. 
We then characterize the topological structure of the emergent Beltrami flow by numerically computing the entanglement statistics of vortex lines in \S~\ref{sec:linking_stats}.
To validate this characterization, we observe that the mean linking number between two vortex lines converges to the asymptotic results due to \citet{arnold_asymptotic_1974}. 
Beyond this result, the full distribution of linking numbers is well described by a k-Gamma distribution, in agreement with an entropic argument previously encountered in granular and living matter for distributions of one-sided random variables with constraints on the mean \citep{aste_emergence_2008,atia_geometric_2018,day_cellular_2022}. This statistical argument is detailed in \S~\ref{sec:linking_dist}.

\section{Chiral symmetry breaking in linearly forced active flows}
\label{sec:chiral}

\begin{figure}
\centerline{\includegraphics{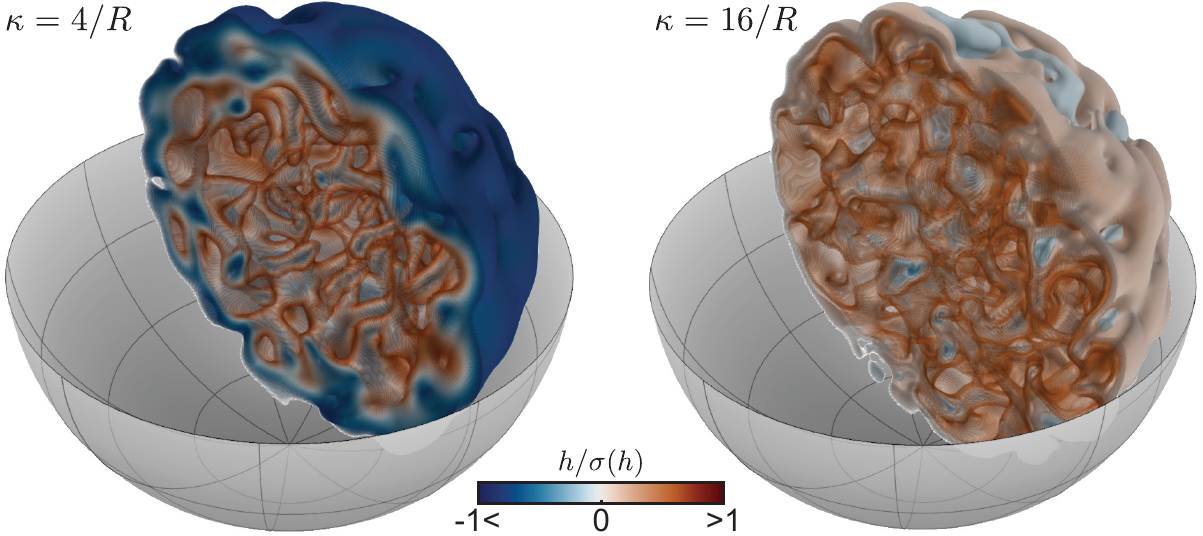}}
\caption{
    Simulations of the active GNS model Eq.~\eqref{eq:GNS_all} confined in the ball demonstrate spontaneous parity-breaking and helical flows. 
    Structures are more pronounced for a narrow spectral energy injection bandwidth ($\kappa=4/R$, left) than a wider one ($\kappa=16/R$, right). 
    Iso-value surfaces of the helicity density $h=\bm{u}\bm{\cdot} \boldsymbol{\omega}$ are shown, rescaled by the standard deviation $\sigma(h)$. 
    Regions of positive helicity (red) dominate in the bulk and negative regions (blue) are mostly present near the boundary. 
    The vortex scale parameter is $\Lambda = R/8$ in both simulations, and time units are chosen such that $\tau=1$.
}
\label{fig:fig_3d}
\end{figure}

The unbounded GNS equations admit exact chiral solutions. 
These solutions are Beltrami flows, in which the vorticity is colinear with the velocity \citep{slomka_spontaneous_2017}. 
More specifically, the exact GNS solutions have a velocity field which is an eigenfunction of the curl operator \begin{equation}
    \boldsymbol{\omega} =\nabla \times \bm{u} = \lambda \bm{u}, \label{eq:beltrami_def}
\end{equation}
with $\lambda$ the eigenvalue corresponding to the characteristic wavenumber of the mode; such solutions are sometimes called Trkalian flows~\citep{lakhtakia_viktor_1994}.
\rev{The initial linear instability drives modes with $\lambda \approx \pi/\Lambda$ and both positive and negative helicities. Triadic interactions break the symmetry and spontaneously select an overall handedness for the flow \citep{slomka_nature_2018}.
Simulations in periodic domains that start from initially random small velocity fields therefore spontaneously produce statistically stationary chiral flows.}
It remains, however, to observe whether such solutions can robustly manifest themselves in the presence of boundaries.

Using the spectral solver Dedalus \citep{burns_dedalus_2020}, we simulate a confined GNS flow inside the three-dimensional ball of radius $R$ (figure~\ref{fig:fig_3d}). 
Starting from an initially small random flow, we evolve the GNS equations (eqs.~\eqref{eq:GNS_all}) subject to the boundary conditions:
\begin{subequations}
\begin{align}
    \bm{u}(R,t) & = 0 \label{eq:BC_1}\\
    \bm{n}\bm{\cdot}\nabla\bm{u}(R,t) & = 0 \label{eq:BC_2}\\
    \bm{n}\bm{\cdot}\bm{n}\bm{\cdot}\nabla\nabla\bm{u}(R,t) & = 0 \label{eq:BC_3}
\end{align}\label{eq:GNS_BCs}%
\end{subequations}
\rev{The no-slip condition here ensures there is no normal vorticity at the boundary (a necessary condition for Arnold's theorem), and the higher order terms are chosen to simply suppress shear at the boundary, but other choices are possible \citep{slomka_geometry_2017}.}
Throughout, we set $\Lambda = R/8$ and adopt time units such that $\tau=1$. 
To simulate the GNS equations~\eqref{eq:GNS_all} with boundary conditions~\eqref{eq:GNS_BCs}, we discretize the ball along the $(r,\theta, \phi)$ coordinates using $(N_r, N_\theta, N_\phi) = (64, 64, 128)$ grid points. 
Time stepping is done using a 3rd-order 4-stage implicit-explicit Runge-Kutta scheme \citep{ascher_implicit-explicit_1997}. 
Additional details of the numerical implementation \rev{and initial condition construction} are summarized in Appendix~\ref{appA}.

\begin{figure}
\centerline{\includegraphics{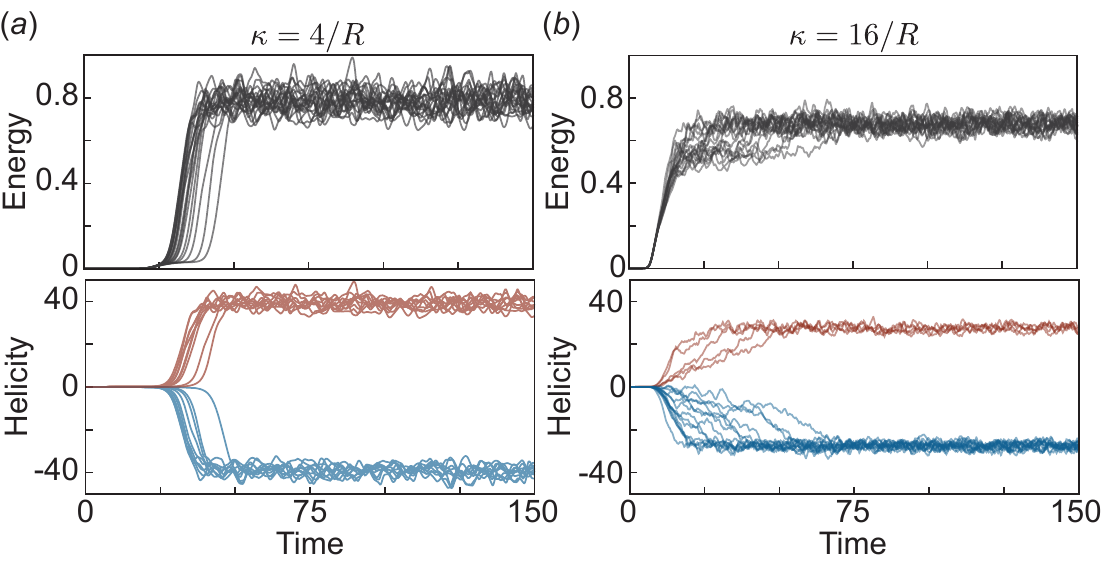}}
\caption{
    Spontaneous chiral flows in confined GNS: Energy and total helicity as a function of time for different initial conditions. 
    \textit{(a)} 20 simulations with $\Lambda = R/8, \kappa=4/R$.
    \textit{(b)} 19 simulations with $\Lambda = R/8, \kappa=16/R$. 
    In both (a) and (b), the initial conditions spontaneously break parity symmetry, and $\tau=1$ sets the time unit.
}
\label{fig:fig_symbreak}
\end{figure}

As has been observed in periodic domains \citep{slomka_spontaneous_2017}, after an initial transient, the GNS dynamics lead to the \rev{spontaneous} emergence of vortical flows which saturate at a finite energy \rev{$E(t) = (1/2)\int_V \mathrm{d}^3\bm{x}\, \bm{u}^2(\bm{x},t)$} (figures~\ref{fig:fig_3d} and \ref{fig:fig_symbreak}). 
The asymmetry between positive and negative helicity density regions suggests that the emergent flow violates parity invariance (figure~\ref{fig:fig_3d}): \rev{while the GNS equations -- including the chosen boundary conditions -- are invariant under the transformation $\bm{x} \rightarrow -\bm{x}$, solutions with a non-zero total helicity are not invariant under this transformation.} 
To quantify the extent of this parity-symmetry breaking, we compute the helicity of the flow $H(t)$ through its integral definition. 
We find that $H(t)$, \rev{like the energy $E(t)$, starts at a small value and grows until saturating at a much larger amplitude. But unlike the energy, the steady-state helicity has a sign that is determined by the random initial condition (figure~\ref{fig:fig_symbreak}).
It is interesting to note the variability in transient behaviour between samples, with the presence of dynamics on multiple timescales.} 
This phenomenon is reminiscent of mode competition in nonlinear systems such as multimode lasers \citep{hodges_turn-transient_1997} \rev{and population dynamics \citep{hastings_transient_2018, morozov_long_2020}}. 

\begin{figure}
\centerline{\includegraphics{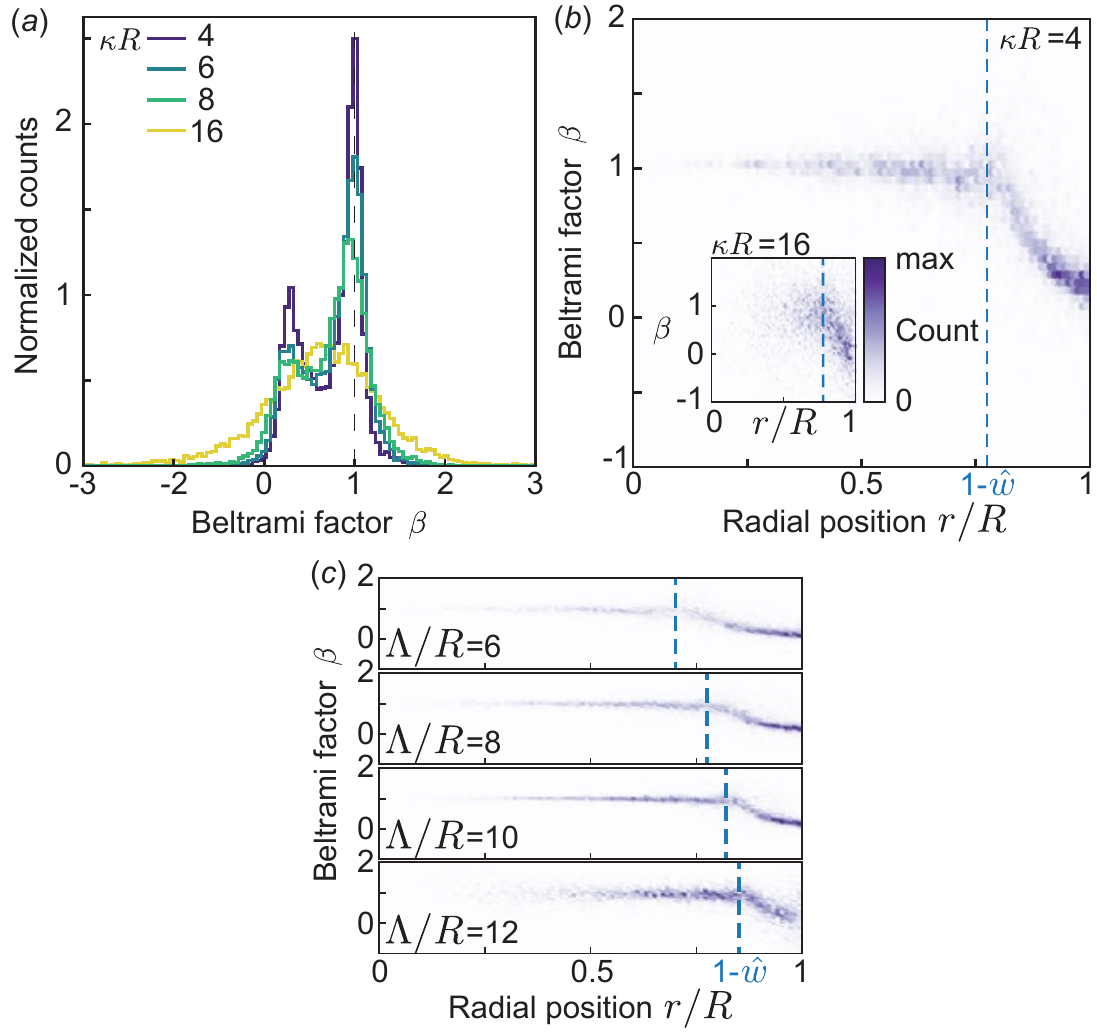}}
\caption{
    GNS spontaneously produces bulk quasi-Beltrami flows for narrow unstable bandwidth: \textit{(a)} The Beltrami factor $\beta = \bm{u}\bm{\cdot} \boldsymbol{\omega}/ (\lambda |\bm{u}^2|)$ with $\lambda = \pi/\Lambda$ is increasingly peaked for systems with narrower spectral bandwidth $\kappa$. 
    The depicted flows all have positive helicity, but opposite parity solutions appear with equal probability. 
    \textit{(b)} Two-dimensional histogram of the Beltrami factor against the radial position, revealing an approximate Beltrami flow in the bulk of the ball with adjustments \rev{in a boundary layer of relative thickness $\hat{w} = 6^{1/3}\Lambda/R$.}
    At higher bandwidths (inset), more modes are excited and the Beltrami factor is less clustered. 
    $\Lambda = R/8$ in all simulations of \textit{(a)} and \textit{(b)}.
    \rev{\textit{(c)} Beltrami factor for simulations with $\kappa R=4$ and varying $\Lambda$ support the expected boundary layer scaling.} 
    The histograms in \textit{(b)} and \textit{(c)} are constructed from $10^4$ uniformly random sample points. 
    All values are in simulation units where $R=1, \tau = 1$.
}
\label{fig:fig_beltrami}
\end{figure}

What is the structure of these emergent chiral solutions? 
To compare our flows to the expected Beltrami solutions in the bulk, we compute a `Beltrami factor' $\beta = \bm{u}\bm{\cdot} \boldsymbol{\omega}/ (\lambda |\bm{u}|^2)$ with $\lambda = \pi/\Lambda$. 
If the flow followed the structure of the periodic solutions \eqref{eq:beltrami_def}, we would expect this measure to be peaked around $\pm1$ in the bulk, with $+1$ corresponding to positive helicity solutions. 
Indeed, as we decrease $\kappa$ and fewer modes are excited, $\beta$ peaks around $1$, with the notable appearance of secondary peaks (figure~\ref{fig:fig_beltrami}a). 
Those new peaks can be simply explained: in confined domains, solutions to Eq.~\eqref{eq:beltrami_def} cannot also be solutions to the GNS equations \eqref{eq:GNS_all} subject to the chosen boundary conditions \eqref{eq:GNS_BCs}. 
This frustration leads to the appearance of a boundary layer (as can be seen in figure~\ref{fig:fig_3d} for $\kappa=4/R$), and indeed $\beta$ is peaked around unity in the bulk of the sphere (figure~\ref{fig:fig_beltrami}b). 
\rev{By Taylor expanding near $r=R$ to thrid order with the chosen boundary conditions \eqref{eq:GNS_BCs}, and matching to the typical third derivative in the bulk $\sim U/\Lambda^3$, the characteristic boundary layer scale is expected to be $w = 6^{1/3}\Lambda \approx 1.8 \Lambda$, which matches well with our simulations (figure~\ref{fig:fig_beltrami}bc).}

The generalized Navier-Stokes equation hence spontaneously generate quasi-Beltrami flows in the bulk for narrow energy injection bandwidths. 
Why do GNS solutions converge to such flows? 
Chiral symmetry breaking has been explained in previous work by noticing that the advection term selects for chiral solutions in the bulk in the presence of energy injection by linear instability \citep{slomka_nature_2018}. 
This selection effect is theorized to be more pronounced as the energy injection bandwidth $\kappa$ narrows, in accordance with our numerical observations, as both the absolute value of the helicity and $\beta$ decrease with larger bandwidths (figure~\ref{fig:fig_symbreak}b,\ref{fig:fig_beltrami}). 
As Beltrami flows minimize enstrophy $\mathcal{E} = (1/2)\int \mathrm{d}^3\bm{x} \,\boldsymbol{\omega}^2$ at fixed helicity \citep{woltjer_theorem_1958} \rev{and are stable solutions to the Euler equations}, one might speculate that \rev{in our effectively inviscid flows \citep{lopez_turning_2015}} the selection of helical modes naturally leads to such Beltrami flows away from boundaries \citep{slomka_spontaneous_2017}.
\rev{While out-of-equilibrium dynamics do not necessarily follow any extremization principle, we note that this property of Beltrami flows is purely geometric and does not depend on the nature of the flow}.
It would thus be interesting to see under what conditions other helical turbulence generation mechanisms also produce Beltrami flows, and whether this regime could be realized in microfluidic experiments using semi-dense bacterial suspensions \citep{wioland_confinement_2013} or other biological or synthetic active matter.

\section{Quantifying chiral symmetry breaking through vortex linking statistics}
\label{sec:linking_stats}

In the previous section, we showed that the GNS flow in the ball spontaneously produces quasi-Beltrami flows with non-zero helicity.
To connect the helicity of the flow to the linking statistics of vortex lines we follow an approach inspired by theorems by \citet{moffatt_degree_1969} and \citet{arnold_asymptotic_1974}.

Both theorems are concerned with the instantaneous entanglement of vortex lines, which are defined as the streamlines (in the mathematical sense) of the vorticity field, satisfying the differential equation
\begin{align}
    \frac{\mathrm{d}\bm{r}}{\mathrm{d}\Tilde{s}} & = \boldsymbol{\omega}(\bm{r}(\Tilde{s}))
    \label{eq:vortex_line}
\end{align}
where $\Tilde{s}$ is a parameter with units of length $\times$ time. 
Here and in what follows, this integration is always performed at a fixed simulation timepoint $t$, considering the vorticity field $\boldsymbol{\omega}$ as a `frozen-in' structure at time $t$. 
To avoid numerical issues due to varying magnitude of $\boldsymbol{\omega}$, we consider the equivalent differential equation 
\begin{align}
    \frac{\mathrm{d}\bm{r}}{\mathrm{d}s} & = \frac{ \boldsymbol{\omega}}{| \boldsymbol{\omega}|}(\bm{r}(s))
    \label{eq:vortex_line_2}
\end{align}
now re-parameterized such that a vortex line integrated from an initial position $\bm{r}(0)=\bm{r}_0$ over $s\in [0, L]$ has length $L$. 
We numerically integrate Eq.~\eqref{eq:vortex_line_2} by using a linear interpolation of the field $\boldsymbol{\omega}$ between the quadrature nodes of the spectral direct numerical simulation.

\begin{figure}
\centerline{\includegraphics{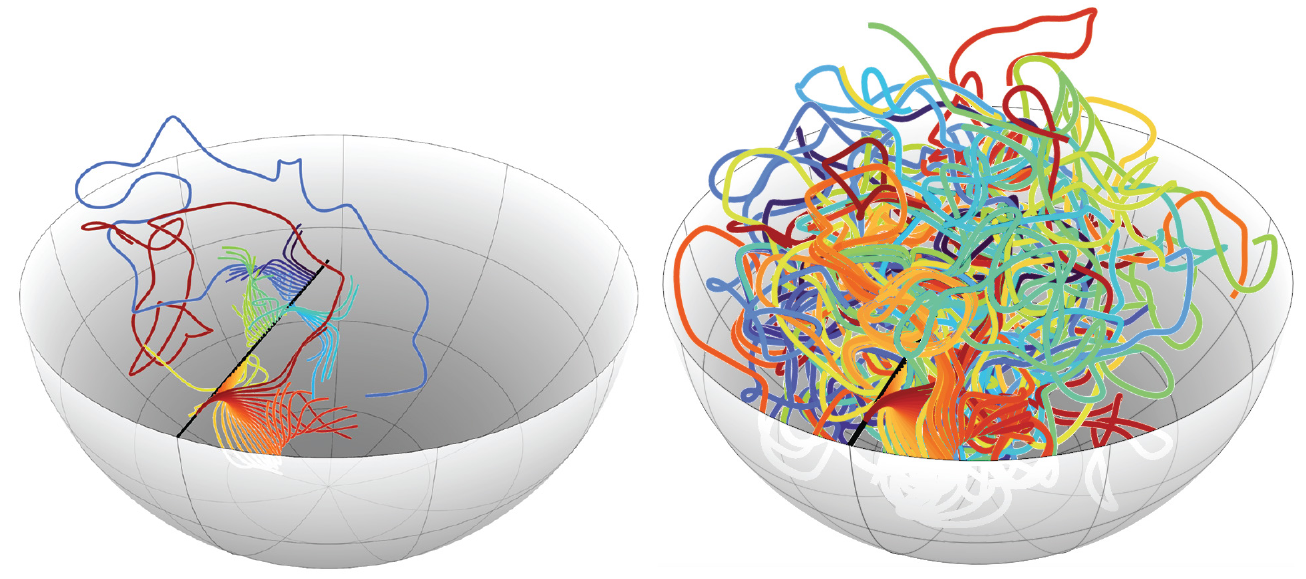}}
\caption{
    Vortex lines, corresponding to integral curves of the vorticity field at fixed time $t$ as defined in Eq.~\eqref{eq:vortex_line}, starting at evenly spaced intervals along the black line, rapidly diverge and tangle around each other. 
    Left: lines of length $L\approx 0.3R$; the two longer lines have $L\approx 0.9R$. 
    Colour indicates starting position. 
    Right: the same lines extended to $L=5R$. 
    The vorticity field is from the statistically stationary turbulent state with $\Lambda = R/8$, $\kappa=4/R$.}
\label{fig:fig_noodle}
\end{figure}

Vortex lines are experimentally accessible flow structures in the limit of localized vorticity, as buoyant particles such as bubbles in water are attracted to regions of high vorticity while heavier particles are expelled away from them. 
This makes vortex lines readily visible in flows where vorticity is confined to tube-like regions \citep{kleckner_creation_2013, durham_turbulence_2013}.
However, vortex line geometry is often complex even in the presence of relatively simple flow fields, as illustrated by the chaotic field lines in the Arnold-Beltrami-Childress (ABC) flow \citep{dombre_chaotic_1986, qin_kind_2023}. 
Integration of vortex lines in our helical flows indeed lead to erratic trajectories, where initially close vortex lines rapidly diverge and tangle around each other~(figure~\ref{fig:fig_noodle}).

\begin{figure}
\centerline{\includegraphics{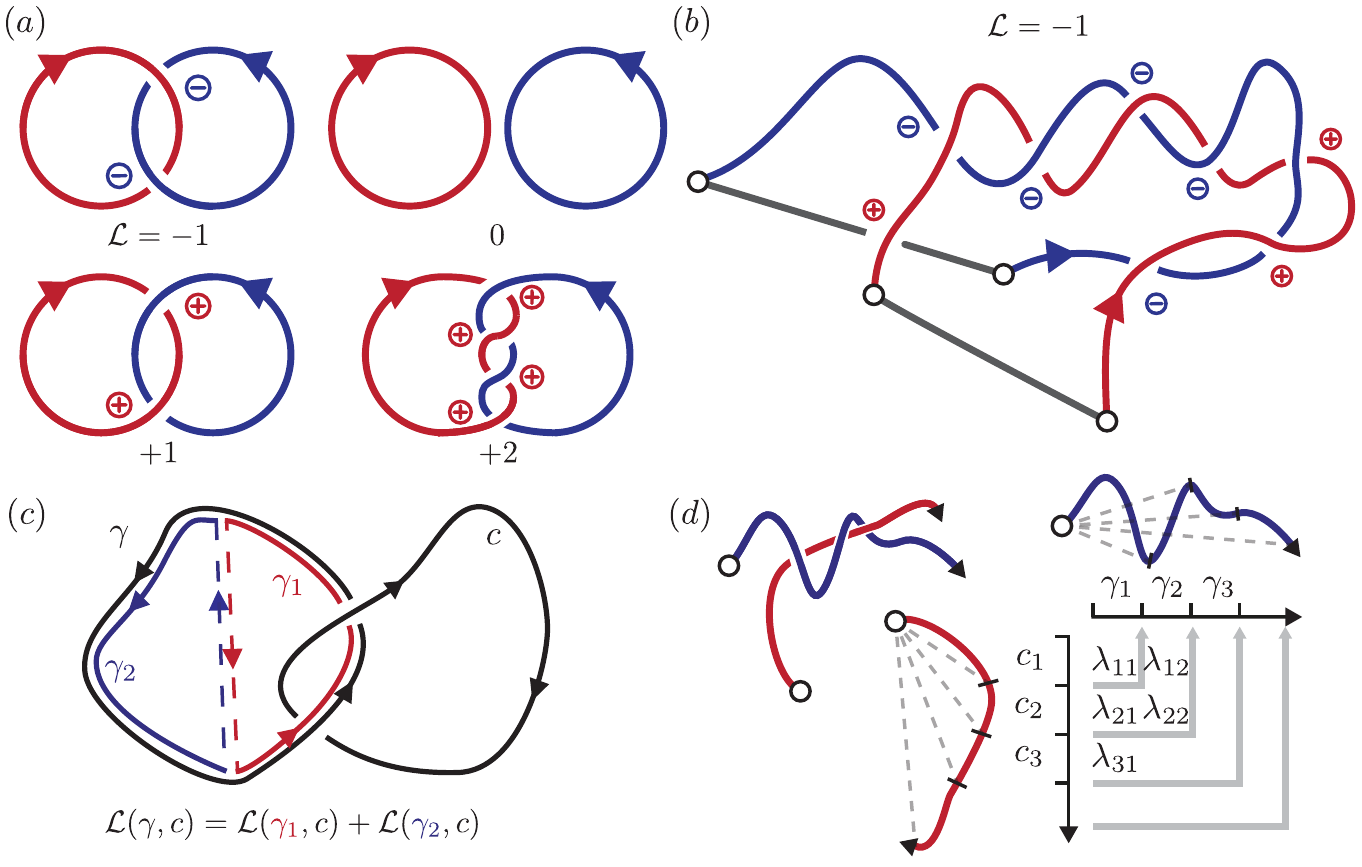}}
\caption{
    Linking numbers are topological invariants measuring the entanglement of oriented closed curves: (\textit{a}) Examples configurations with their linking numbers. 
    Positive and negative crossings are illustrated by $\oplus$ and $\ominus$. 
    (\textit{b}) Linking number of two example open curves closed using the scheme used to apply Arnold's theorem. 
    (\textit{c}) Illustration of the additive property of linking numbers with respect to curve concatenation. 
    (\textit{d}) The additive property can be used to compute linking numbers as sums of contributions from subsegments of two curves.
}
\label{fig:fig_linking}
\end{figure}

To measure pair-wise entanglement of vortex lines, we use the linking number between two oriented closed curves $\gamma_1$ and $\gamma_2$ defined by Gauss' integral formula \citep{qu_fast_2021}
\begin{equation}
    \mathcal{L}(\gamma_1, \gamma_2) = \frac{1}{4\pi}\oint_{\gamma_1} \oint_{\gamma_2} \frac{(\bm{r}_1 - \bm{r}_2)}{\,\,\,|\bm{r}_1 - \bm{r}_2|^3}\bm{\cdot} (\mathrm{d}\bm{r}_1 \times \mathrm{d}\bm{r}_2).\label{eq:Lk_def}
\end{equation}
This integer-valued quantity counts the signed number of times one curves winds around the other. 
The linking number possesses notable properties: if one curve is reversed, the linking number flips sign, and it is symmetric since $\mathcal{L}(\gamma_1, \gamma_2) = \mathcal{L}(\gamma_2, \gamma_1)$. 
Most importantly, $\mathcal{L}$ is invariant under continuous deformation of the curves. 
The linking number is therefore a topological invariant playing a central role in the study of knots and linked curves \citep{kauffman_knots_1995,vologodskii_conformational_1994,adams_topological_2019}.
An equivalent viewpoint is to define $\mathcal{L}$ as the sum of half-integer contributions from each crossing of the curves under a planar projection, with the sign depending on the relative orientation of the curves (figure~\ref{fig:fig_linking}a). 
As an invariant, $\mathcal{L}$ is independent of the choice of the projection plane.

While the definition of $\mathcal{L}$ in Eq.~\eqref{eq:Lk_def} calls for closed curves, vortex lines in complex flows are not closed in general. 
To leverage the connections between linking numbers and helicity, we thus have to consider slightly modified vortex lines. 
Consider two open vortex lines $\gamma_1$ and $\gamma_2$ of length $L$ starting at points $\bm{x}_1$ and $\bm{x}_2$, respectively. 
We then define the asymptotic linking number $\Lambda_L(\bm{x}_1,\bm{x}_2)$ between $\gamma_1$ and $\gamma_2$ as follows. 
We construct the curve $\Tilde{\gamma}_i$ as by closing the curve $\gamma_i$ by a straight segment connecting its start and end points. 
We then note $\Lambda_L(\bm{x}_1,\bm{x}_2)$ the linking number of the two curves $\Tilde{\gamma}_1$, $\Tilde{\gamma}_2$ (figure~\ref{fig:fig_linking}b) normalized by $T_i$ and $T_j$
\begin{equation}
    \Lambda_L(\bm{x}_1,\bm{x}_2) = \frac{1}{T_1 T_2}\mathcal{L}(\Tilde{\gamma}_1, \Tilde{\gamma}_2),
\end{equation}
where $T_i = \int_{\gamma_i} \mathrm{d}s/|\omega(\bm{r}(s))|$ is the integrated inverse circulation, which has units of length $\times$ time; the product $T_iT_j$ is the normalization factor of the asymptotic linking number.

With these preliminary definitions, \citet{arnold_asymptotic_1974} provides a connection between the helicity of an incompressible flow and the asymptotic linking number of vortex lines via the asymptotic equality:
\begin{equation}
    H=\int_V \mathrm{d}^3\bm{x} \,(\bm{u}\bm{\cdot} \boldsymbol{\omega})(\bm{x}) = \frac{1}{V^2}\iint_{V\times V}\mathrm{d}^3\bm{x}_1\mathrm{d}^3\bm{x}_2 \lim_{L\rightarrow+\infty}\Lambda_L(\bm{x}_1, \bm{x}_2). \label{eq:Arnold_helicity}
\end{equation}
This equality between the helicity and the volume averages of the asymptotic linking numbers is valid on any simply connected domain for any incompressible velocity field, as long as the normal component of the vorticity vanishes at the boundary ($\bm{n}\bm{\cdot} [\nabla \times \bm{u}] = 0$). 
Since our simulations have a no-slip boundary condition $\bm{u}=0$, by Stokes' theorem the vorticity flux through any arbitrary loop drawn on the boundary cancels, and we can apply Eq.~\eqref{eq:Arnold_helicity}.

\begin{figure}
\begin{centering}
\centerline{\includegraphics{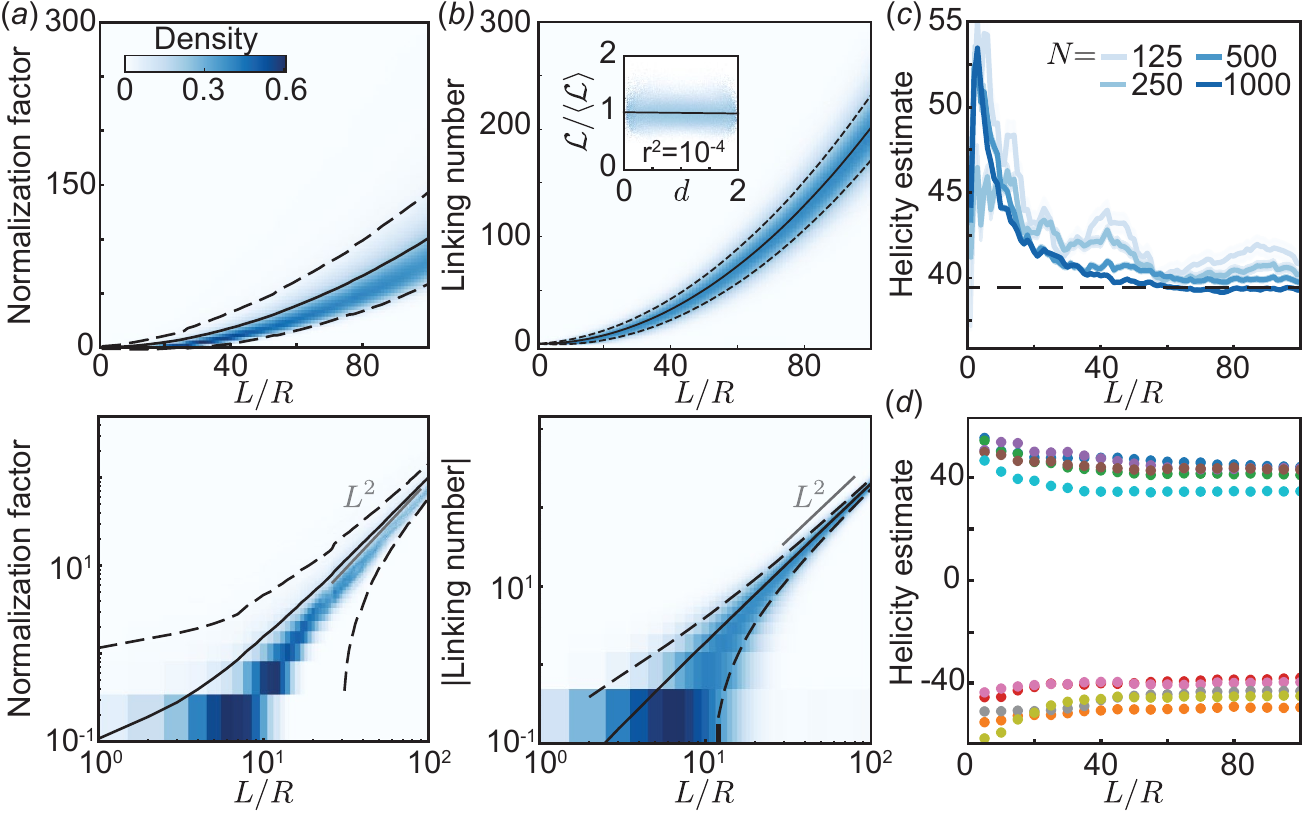}}
\caption{
    The average linking number converges to the total helicity as vortex line length is increased: \textit{(a)} Heatmap of the normalization factor $T_iT_j$ as a function of vortex line length $L$ averaged over $N = 1000$ vortex lines, in linear (top) and log scale (bottom). 
    Continuous lines indicate mean, dashed lines standard deviation.
    Notice that the mean sits above the mode of the distribution, as the normalization factor distribution has a long tail due to interactions with the boundaries. 
    \textit{(b)} Heatmap of the pair-wise linking number $\mathcal{L}$ as a function of vortex line length $L$, in linear (top) and log scale (bottom) for a positive helicity flow. 
    For better visualization, each column in (a), (b) is normalized to the column-wise standard deviation. 
    Results are averaged over $N = 1000$ vortex lines and their $N(N-1)/2 = 5\times10^5$ pairs. 
    The logarithmic plot shows an asymptotic scaling of $\mathcal{L} \sim L^2$. 
    Inset shows the set of linking numbers $\mathcal{L}$ for $L=100R$ as a function of the distance $d$ between the vortex lines starting points. 
    There is no correlation between $\mathcal{L}$ and $d$. 
    \textit{(c)} Helicity estimate as a function of vortex line length for increasing number of vortex lines. 
    \textit{(d)} Helicity estimate for various samples (different colours) computed from $N=500$ vortex lines as a function of $L$. 
    Remarkably, even short lines capture the helicity sign, suggesting that even limited observations of vortex lines could be used to detect chiral symmetry breaking in experiments.
}
\label{fig:fig_helicity_cv}
\end{centering}
\end{figure}

To characterize the topology of the active Beltrami flow, we hence construct a statistical ensemble of linking numbers using the numerical solutions of Eq.~\eqref{eq:vortex_line_2}, verifying our approach against the expectation of Eq.~\eqref{eq:Arnold_helicity} for the mean of this distribution.

As a first step towards an estimate of the helicity using Arnold's theorem, we integrate $N$ vortex lines of length $L$ and then close them as in figure~\ref{fig:fig_linking}b. 
To compute the linking number between two such closed vortex lines, which are numerically represented as a set of connected line segments with lengths $\delta<10^{-2}R$, a naive discretization of the Gauss integral of Eq.~\eqref{eq:Lk_def} between two lines of length $L$ would require $O(L^2/\delta^2)$ operations. 
This quadratic scaling leads to a prohibitive computational cost for large linking numbers. 
However, $N$-body simulations often require the evaluation of integrands which decay as $\propto 1/r^2$, as is the case for Eq.~\eqref{eq:Lk_def}. 
For this class of functions, one can leverage Barnes-Hut and fast multipole-type methods to bring down the algorithmic complexity of the linking number computation to $O((L/\delta)\log (L/\delta))$. 
\citet{qu_fast_2021} designed and implemented such methods for computing linking numbers, along with topology-preserving curve simplification algorithms in a publicly available C++ package. 
We have built and released a Python wrapper for their code (See Data availability statement).

To monitor convergence of the mean of the asymptotic linking numbers to $H$ as a function of line length $L$, we can exploit the linearity of the curve integrals in Eq.~\eqref{eq:Lk_def}. 
By decomposing the linking number into contribution from subloops, we can avoid recomputing linking numbers from scratch \citep{moffatt_degree_1969}. 
Let $\gamma_1 \gamma_2$ denote the concatenation of the oriented closed curves $\gamma_1$ and $\gamma_2$ sharing a start and end point, then we have $\mathcal{L}(\gamma_1\gamma_2,\gamma_3) = \mathcal{L}(\gamma_1,\gamma_3) + \mathcal{L}(\gamma_2,\gamma_3)$ (figure~\ref{fig:fig_linking}c). 
Contributions from subloops can then be summed up to recover the full linking number of longer curves (figure~\ref{fig:fig_linking}d).

With these ingredients in hand, we can compute the linking number distribution and construct an `Arnold estimate' of the helicity, by constructing a Monte-Carlo approximation to the integral in the right-hang side of Eq.~\eqref{eq:Arnold_helicity} as follows:
\begin{enumerate}
    \item Sample $N$ initial points $\{\bm{x}_i\}$ uniformly in the domain.
    \item Integrate vortex lines and their inverse circulation for a length $L$. Closing the vortex lines by a line segment, we obtain a set of $\{g(\bm{x}_i), T_i\}$.
    \item Compute the $N(N-1)/2$ distinct linking numbers $\lambda_{ij} := \mathcal{L}\left(g(\bm{x}_i),g(\bm{x}_j)\right)$ using fast methods \citep{qu_fast_2021}.
    \item Normalize the linking numbers to obtain the approximations $\Lambda_L$ and average those contributions to estimate the average helicity :
    \begin{equation}
        \hat{H} = \frac{2}{N(N-1)}\sum_{0<i<j\leq N} \frac{1}{T_i T_j}\lambda_{ij}. \label{eq:Hhat}
    \end{equation}
\end{enumerate}
Applying this program, we study the convergence of the estimate $\hat{H}$ to $H$ as a function of $N$ and $L$ for a snapshot of a given simulation at a fixed timepoint once the GNS has entered the statistically stationary state.

Integrating vortex lines, we find normalization factors $T_iT_j$ increasing as $\sim L^2$, as can be expected for long vortex lines with $L\gg R$ which traverse the entire domain such that $T_i \approx L/\langle |\boldsymbol{\omega}|\rangle$, with $\langle . \rangle$ denoting the volume average (figure~\ref{fig:fig_helicity_cv}a). 
Note that boundaries are irrelevant to the computation: as $T_i$ diverges to $+\infty$ when $\bm{u}$ vanishes, vortex lines intersecting the boundary layer have very large normalization factors. 
The contributions from such `boundary' vortex lines are hence suppressed from the estimate Eq.~\eqref{eq:Hhat}, and the mean of $T_iT_j$ is larger than its mode. 
Once the vortex lines are computed, we compute our ensemble of linking numbers; consistent with the implication of Eq.~\eqref{eq:Hhat} that $\Lambda_L$ must converge to a finite value, our computed linking numbers have their average value scaling with with $L^2$ (figure~\ref{fig:fig_helicity_cv}b).
It is interesting to note that vortex lines longer than a few domain radius will almost certainly link with other vortex lines with the sign of the total helicity. 
Additionally, the linking number between vortex lines is independent of vortex line starting position; this is consistent with the picture that vortex line integration is chaotic (figure~\ref{fig:fig_helicity_cv}b, inset). 

Combining linking numbers and normalization factors to compute $\hat{H}$, we find rapid convergence with length and number: $N=125$ vortex lines of length $\sim30R$ give an estimate accurate within 10\%. 
Remarkably, even short vortex lines lead to the correct sign and order of magnitude of the helicity, across tested samples (figure~\ref{fig:fig_helicity_cv}d). 

\begin{figure}
\centerline{\includegraphics{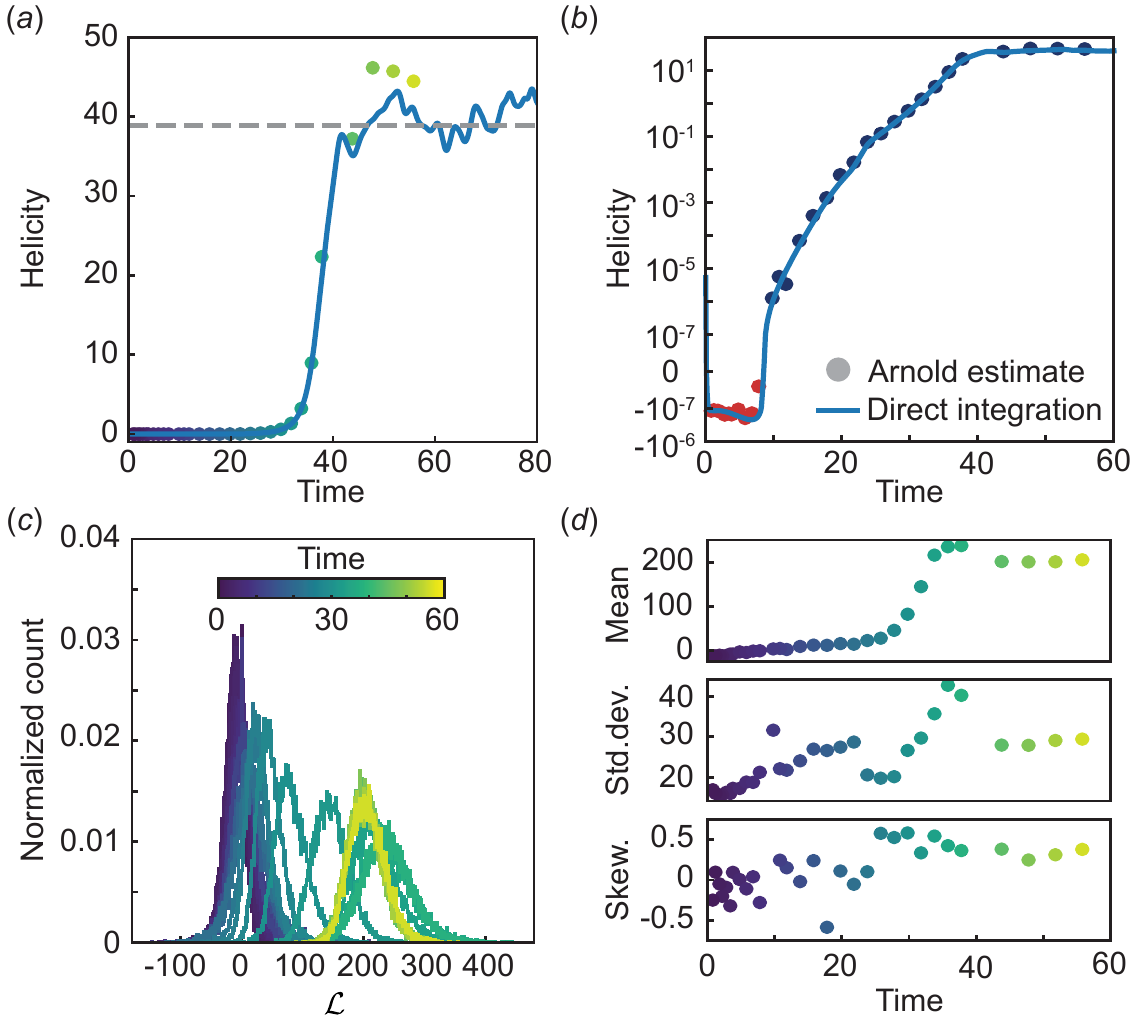}}
\centering 
\caption{
    Tracking chiral symmetry-breaking through vorticity linking:
    \textit{(a)} Direct integration of the helicity density (solid line) matches the Arnold estimate of the helicity by vortex linking number statistics (solid circles). 
    Dashed line indicates the average helicity at steady-state. 
    \textit{(b)} The Arnold estimate is accurate even at short times, and notably captures the helicity sign change, with red dots denoting negative values and blue dots indicating positive values. 
    Y-axis is linear between $[-10^{-7}, 10^{-7}]$ and logarithmic elsewhere. 
    \textit{(c)} Time evolution of the probability distribution of pair-wise linking numbers. 
    Colour indicates time. 
    \textit{(d)} Mean, standard deviation and moment coefficient of skewness of the linking number distribution as a function of time. 
    The final non-zero mean reflects chirality of flow, while increasing skewness indicates the departure from Gaussian statistics. 
    We note that at time $10$, an outlier point with skewness $40$ is not shown. 
    Results are shown for $N= 150$ vortex lines of length $L=100R$. 
    Colour of scatter points in \textit{(a)}, \textit{(c)}, and \textit{(d)} all follow the colour map of \textit{(c)}.
}
\label{fig:fig_topo_time}
\end{figure}

Settling on $N=150, L=100R$ to construct our Arnold estimate $\hat{H}$, we run the above algorithm at various time points to monitor the time-evolution of vorticity linking (figure~\ref{fig:fig_topo_time}). 
The Arnold estimate is an accurate estimate of the helicity during the initial period, linear instability, and saturation phase, with an approximately constant relative error (figure~\ref{fig:fig_topo_time}a,b). 
In line with our previous observation that even short vortex lines led to the correct helicity sign, $\hat{H}$ has the correct sign even when the helicity magnitude is close to zero. 
This raises the interesting possibility of tracking chiral symmetry breaking in experiments using tracer particles to uncover vortex lines, especially at low tracer densities that would make standard velocity reconstruction methods challenging.

The dynamics of vortex lines are key to understanding the emergence of fine structures in turbulence \rev{through their connection to helicity as an inviscid invariant} \citep{scheeler_complete_2017,mckeown_turbulence_2020,matsuzawa_creation_2023}. 
\rev{In active flows, however, helicity can be created and destroyed.
Our techniques allow the characterization of the statistically-averaged topology of the flow as it evolves to a statistically stationary state with nonzero total helicity.}
Here, our construction of $\hat{H}$ provides us with the full distribution $p(\mathcal{L})$ of linking numbers as a function of time, beyond Arnold's result on the mean degree of linkage of vortex lines (figure~\ref{fig:fig_topo_time}c). 
Studying the moments of this distribution, one finds that as expected, the mean linking number goes from $0$ to finite non-zero number reflecting the emergence of chiral flows (figure~\ref{fig:fig_topo_time}d). 
The behaviour of the standard deviation and skewness are however non-trivial: the standard deviation shows large fluctuations during the instability growth phase, while the distribution displays non-zero skewness at late times. 
In the next section, we will use general constraints on the linking number distribution to rationalize its statistics at steady-state.

\section{The distribution of linking numbers obeys a maximum-entropy law} 
\label{sec:linking_dist}

The construction from the previous part allows us to obtain the distribution of pair-wise linking numbers of $N$ vortex lines of length $L$, which contains several notable features at steady-state. 
First, for the strongly chiral flows considered here, all pairs of sufficiently long vortex lines are linked with probability $1$. 
Second, the mean $\langle \mathcal{L}\rangle$ is approximately constrained by Arnold's theorem to be equal to the flow's total helicity. 
Third, we find no correlation between linking numbers and vortex line starting points for sufficiently long vortex lines, suggesting `chaotic tangling' and a notion of ergodicity in the system, with two randomly selected vortex lines eventually capturing the global helicity as their lengths tend to infinity. 
Together, these features suggest that in this geometrically and topologically complex system, statistical principles could explain the observed linking distribution.

Maximal-entropy reasoning has been successfully applied to explain the packing statistics of confined granular and living matter \citep{edwards_theory_1989,bi_statistical_2015, day_cellular_2022,atia_geometric_2018} and topological defect distributions in two-dimensional turbulence \citep{eyink_onsager_2006, giomi_geometry_2015}. 
As these problems naturally share features with our system of confined topological defects, the maximal-entropy method is a viable candidate to explain linking number statistics.

To apply a maximum-entropy approach, we translate the above observations into constraints that the distribution of linking number must plausibly satisfy. 
The first observation implies that for a given length $L$, the linking numbers must be bounded from below $\mathcal{L} \geq \mathcal{L}_\text{min}$ for a positive helicity flow; in negative helicity flows, $\mathcal{L} \leq \mathcal{L}_\text{max}$. 
The second observation implies that the sum of linking numbers must be approximately equal to the flow helicity by Eq.~\eqref{eq:Arnold_helicity}
\begin{equation}
    \sum_{i,j} \mathcal{L}_{ij} \approx \left(\frac{VNL}{\langle | \boldsymbol{\omega}|\rangle}\right)^2 H \equiv \Bar{H}. \label{eq:mean_to_H}
\end{equation}
Here, we assumed long enough vortex lines $L \gg R$ such that we can take the flow to be homogeneous and $T_i\approx \langle | \boldsymbol{\omega}|\rangle/L$. 
Numerically, we do observe $\sum_{i,j} \mathcal{L}_{ij}$ to scale with $L^2$ (figure~\ref{fig:fig_helicity_cv}b). 
Finally, the third observation above suggests that there is effectively no correlation between linking numbers, or even perhaps between linking numbers of `long enough' vortex line sub-segments; under this assumption, one can consider the linking number distribution as drawn from an emergent thermodynamic ensemble. 

To proceed, we consider the distribution $p(\mathcal{L})$ which maximizes the Shannon entropy subject to the constraints outlined above. 
To this end, we consider $p(\mathcal{L})$ as the probability of the `macroscopic state' where one vortex line of length $L$ links $\mathcal{L}$ times with another vortex line. 
Many `microscopic states' corresponding to possible vortex line conformations are compatible with such macro-states. 
Following \citet{aste_emergence_2008}, we consider the Shannon entropy $\mathcal{S}$ written as
\begin{equation}
    \mathcal{S} = \sum_{\mathcal{L} \in \mathbb{Z}} - p(\mathcal{L})\log p(\mathcal{L}) + \sum_{\mathcal{L} \in \mathbb{Z}} p(\mathcal{L}) S(\mathcal{L}),
\end{equation}
where $S(\mathcal{L})$ is the entropy of the state with linking $\mathcal{L}$. 
Under the assumption that all microscopic states are equiprobable, $S(\mathcal{L}) = \log\Omega(\mathcal{L})$ with $\Omega(\mathcal{L})$ the number of micro-states with linking $\mathcal{L}$. 
Under the maximum entropy principle, $p(\mathcal{L})$ is given by optimizing the entropy functional $\mathcal{S}$ under the helicity constraint
\begin{equation}
    \frac{1}{N^2}\Bar{H} = \sum_{\mathcal{L} \in \mathbb{Z}} p(\mathcal{L}) \mathcal{L}. \label{eq:hbar_constraint}
\end{equation}
The solution of this optimization problem is given by a Boltzmann-type distribution
\begin{equation}
    p(\mathcal{L}) \propto \Omega(\mathcal{L}) e^{-\mathcal{L}/\chi} \label{eq:Boltzmann}
\end{equation}
with $\chi^{-1}$ the Lagrange multiplier fixing the helicity constraint. 
To fully determine the maximal-entropy distribution, the last step is to compute $\Omega(\mathcal{L})$.

Motivated by our observation that even short vortex lines almost-certainly link with others, we consider dividing a vortex line into $k$ sub-loops of an approximately constant size $\delta L$ which we see as characteristic of the domain, such that $k\approx L/\delta L$. 
A mesoscopic description of the linking of two vortex lines can be given by the linking numbers of each sub-loop of the first vortex line with the entire other line $\{\ell_i\}_{i=1,\ldots,k}$; in the case of a positive helicity flow, each sub-loop must link at least $\ell_\text{min} = \mathcal{L}_\text{min}/k$ times and $\sum_i \ell_i = \mathcal{L}$ with the assumption of mutual independence of the $\ell_i$. 
Then, approximating discrete sums as integrals for large enough linking numbers, $\Omega(\mathcal{L})$ is given by the volume of the simplex
\begin{equation}
    \Omega(\mathcal{L}) = \int_{\ell_\text{min}}^{\mathcal{L}} \mathrm{d}\ell_1 \int_{\ell_\text{min}}^{\mathcal{L}} \mathrm{d}\ell_2 \cdots \int_{\ell_\text{min}}^{\mathcal{L}} \mathrm{d}\ell_k \,\delta( \sum_{i=1}^k \ell_i - \mathcal{L}) = \frac{(\mathcal{L}-\mathcal{L}_\text{min})^{k-1}}{(k-1)!} \label{eq:OmegaVolume}
\end{equation}
where the integration bounds correspond to the case of a positive helicity flow, in which linking numbers are bounded from below. 
Combining Eqs.~\eqref{eq:Boltzmann} and \eqref{eq:OmegaVolume} while eliminating $\chi = (\langle \mathcal{L} \rangle- \mathcal{L}_\text{min})/k$ using Eq.~\eqref{eq:hbar_constraint}, we obtain the k-Gamma distribution
\begin{equation}
    p(\mathcal{L}) = \frac{k^k}{\Gamma(k)} \frac{\left( \mathcal{L} - \mathcal{L}_\text{min} \right)^{k-1}}{\left(\langle \mathcal{L} \rangle- \mathcal{L}_\text{min} \right)^k} \exp{\left(-k \frac{\mathcal{L} - \mathcal{L}_\text{min}}{\langle \mathcal{L} \rangle- \mathcal{L}_\text{min}}\right)} \label{eq:kGamma}
\end{equation}
where $\Gamma$ is the Euler Gamma function. 
The k-Gamma distribution can be understood as a Gamma distribution with shape parameter $k$ for the scaled and shifted random variable $(\mathcal{L} - \mathcal{L}_\text{min})/(\langle \mathcal{L} \rangle- \mathcal{L}_\text{min})$. 
Note that in the case of negative helicity flows, we still obtain a k-Gamma distribution by substituting $\mathcal{L} \leftarrow - \mathcal{L}$ and $\mathcal{L}_\text{min} \leftarrow - \mathcal{L}_\text{max}$. 

\begin{figure}
\centerline{\includegraphics{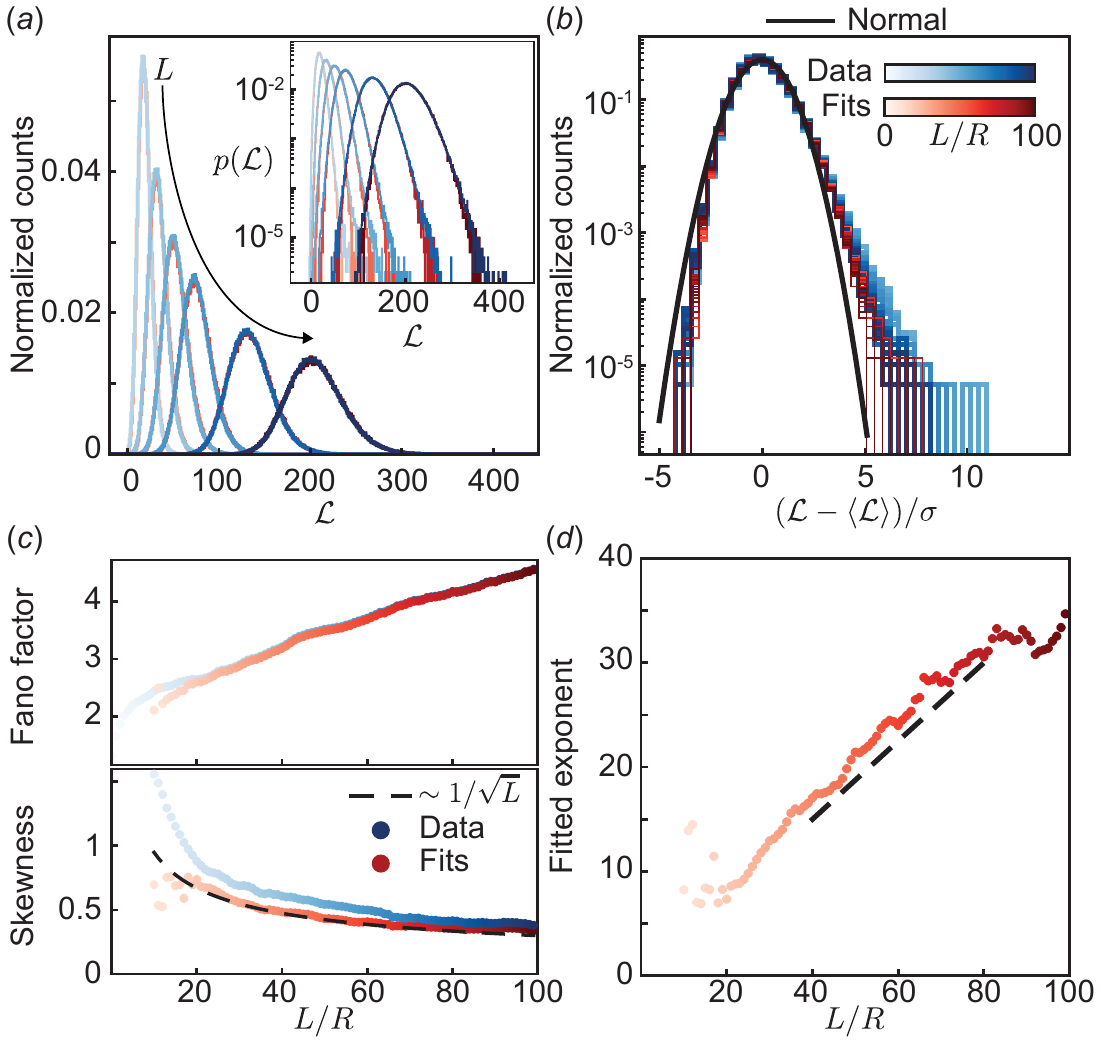}}
\caption{
    The linking number distribution at a fixed time is well-described by a k-Gamma distribution: \textit{(a)} For a flow with positive helicity ($\Lambda = R/8$, $\kappa= 4/R$), the empirical PDF of $\mathcal{L}$ (blue curves) for different vortex line lengths is well-fit by a k-Gamma distribution (red curves), including up to shot noise for long enough vortex lines (log scale inset). 
    Distributions are shown for $L/R=30, 40, 50, 60, 80, 100$, with $N=1000$ vortex lines. 
    \textit{(b)} Plotting centred and scaled distributions for $L/R \geq 50$ highlights the deviation from the normal distribution (black line) and shows the quality of the fit for large $L$. 
    \textit{(c)} The linking distribution has a Fano factor $\sigma^2/\langle \mathcal{L} \rangle > 1$, showing super-Poissonian behaviour that is well captured by the k-Gamma fits (top). 
    The skewness of fits asymptotically matches the data, which scales as $1/\sqrt{L}$, in agreement with the hypothesis of independent increments. 
    \textit{(d)} The fitted exponent $k$ scales linearly with $L$ (dashed lines as visual guide).
}
\label{fig:fig_pdfL}
\end{figure}

 To test the validity of this approach, we fit the probability distribution function in Eq.~\eqref{eq:kGamma} to the linking number distribution obtained from $N=1000$ vortex lines sampled from a positive helicity flow using maximum likelihood estimates of $\mathcal{L}_\text{min}$, $\langle \mathcal{L}\rangle$ and $k$ as a function of the vortex line length $L$ (figure~\ref{fig:fig_pdfL}). 
 We sample $2\times10^4$ points from the fitted distribution and find excellent agreement with the k-Gamma distribution, with $p(\mathcal{L})$ for long ($L=100R$) vortex lines matching the fit down to shot noise. 
 This increase in fit quality with increasing $L$ is consistent with the expectation that our independence and discrete sum approximations become more justified for longer vortex lines (figure~\ref{fig:fig_pdfL}a,b). 

\rev{As expected from Eqs.~\eqref{eq:Arnold_helicity} and~\eqref{eq:mean_to_H}, the fitted value of $\langle \mathcal{L}\rangle$ recovers Arnold's equality.} The fitted k-Gamma distributions also recover the correct scaling of mean and variance of data, notably displaying the super-Poissonian behaviour of the linking distribution as shown by its Fano factor $\sigma^2/\langle \mathcal{L}\rangle >2$, and show the asymptotic behaviour of the skewness (figure~\ref{fig:fig_pdfL}c). 

In previous applications of the k-Gamma distribution, $k$ is fit as a shape parameter determined by the variance $\sigma^2$ of the distribution as $k = (\langle \mathcal{L}\rangle- \mathcal{L}_\text{min})/\sigma^2$ \citep{aste_invariant_2007, aste_emergence_2008, day_cellular_2022,atia_geometric_2018}. 
To obtain parameter estimates as a function of vortex line length, we similarly fit our observed distribution using a maximum likelihood optimization. 
This procedure is agnostic to our arguments leading to Eq.~\eqref{eq:OmegaVolume}, which posits that $k$ is set by the number of independent sub-domains of vortex lines. 
In support of the validity of the decomposition of vortex line linking numbers into contributions from sub-loops, we find that the fitted parameter $k$ linearly increases with the vortex line length $L$ (figure~\ref{fig:fig_pdfL}d). 
\rev{As we find that $k \approx L/(2R)$, we can estimate that topologically-correlated domains have a characteristic size $\delta L \approx 2R$, indicating that this emergent correlation length is set by the diameter of the ball.}

As we consider longer vortex lines, since $k\propto L$ we predict that the linking number distribution must eventually tend to the normal distribution. 
With $k\rightarrow+\infty$, the k-Gamma distribution converges to the Gaussian $\mathcal{N}(\mu=\langle \mathcal{L}\rangle, \sigma^2)$, consistent with figure~\ref{fig:fig_pdfL}b. 
This convergence is also expected of a sum of linking numbers $\ell_i$ from an increasingly large number of statistically independent sub-loops. 
For $\ell_i$ independent and identically distributed, the skewness of the resulting sum is expected to scale as $L^{-1/2}$ when each $\ell_i$ is drawn from a skewed distribution \citep{hall_principles_1992}. 
Further validating the mesoscopic picture that we used to compute $\Omega(\mathcal{L})$, this inverse-square root scaling behaviour is observed in figure~\ref{fig:fig_pdfL}c.

\section{Conclusion}
\label{sec:conclusion}

Building on recent progress in spectral simulation techniques for spherical domains \citep{burns_dedalus_2020}, we simulated a generalized Navier-Stokes (GNS) model for actively driven fluid flow in a confined 3D domain to characterize the topology of the spontaneously forming chiral flow. 
Driven by a generic linear instability \citep{rothman_negative-viscosity_1989,beresnev_model_1993,tribelsky_new_1996}, we find that the GNS system produces an active Beltrami flow in the bulk for narrow energy injection bandwidth, a regime that could likely be realized in microfluidic experiments using semi-dense bacterial \citep{wioland_ferromagnetic_2016} or other microbial suspensions.

Leveraging recently published fast algorithms \citep{qu_fast_2021}, we characterized the topological structure of this spontaneous flow through the pair-wise linking numbers of sampled vortex lines.
We explicitly measured the convergence of the mean vortex line entanglement to the total helicity of the flow, as described asymptotically by Arnold. \rev{Importantly, these results apply to all simply confined solenoidal vector fields with appropriate boundary conditions, making our methodology applicable well beyond active matter models, including high Reynolds number and magnetohydrodynamic flows.}
\rev{In the active flow considered here,} we found that a k-Gamma density describes the linking number distribution well, and propose a maximum-entropy argument to explain this result.

We note that there has been substantial work on understanding the asymptotic behaviour of the winding number of two-dimensional random walks with or without chiral drift, in the presence of repulsion or confinement \citep{spitzer_theorems_1958, drossel_winding_1996, wen_winding_2019}. 
In particular, it is known that the winding number of 2D confined random walks eventually becomes normally distributed for long-enough walks. 
This convergence holds even in the presence of chiral drift, although the presence of drift leads to long-lasting skewness in the winding distribution \citep{drossel_winding_1996, wen_winding_2019}. 
In comparison, the study of the linking number of confined random Brownian walks still presents many open questions \citep{orlandini_statistical_2007}, with to our knowledge no proof of convergence to normality, although results exist for the scaling of moments of drift-free polygonal Brownian walks \citep{panagiotou_linking_2010,marko_scaling_2011}. 
The numerically observed convergence to the normal distribution for the linking number of long confined vortex lines (figure~\ref{fig:fig_pdfL}) suggests connections between winding statistics and the linking numbers of random trajectories. 
\rev{Moreover, given the generic nature of our entropic reasoning, it is possible that the k-Gamma distribution is a universal feature of vortex line entanglement in chiral systems}.

Finally, our analysis shows that even a small number of vortex lines with lengths $L \sim R$ are sufficient to infer both the helicity sign and the helicity value. 
This rapid convergence indicates that observations of vortex line entanglement -- for instance through tracer particles \citep{kleckner_creation_2013} or embedded filaments \citep{kirchenbuechler_direct_2014,du_roure_dynamics_2019} -- could provide reliable measurements of chiral symmetry breaking in 3D experimental flows even with limited data.


\backsection[Acknowledgements]{The authors thank Mehran Kardar for valuable discussions and the MIT SuperCloud and Lincoln Laboratory Supercomputing Center for providing HPC resources that have contributed to the results reported within this paper.}

\backsection[Funding]{
    This work was supported by a MathWorks Science Fellowship (N.R.), 
    NSF Award DMS-1952706 (N.R. and J.D.), 
    Alfred P. Sloan Foundation Award G-2021-16758 (J.D.), 
    the Robert E. Collins Distinguished Scholarship Fund of the MIT Mathematics Department (J.D.), 
    the Swiss National Science Foundation Ambizione Grant PZ00P2\_202188 (J.S.), 
    and NASA HTMS Award 80NSSC20K1280 (K.J.B.).
}

\backsection[Declaration of interests]{The authors report no conflict of interest.}

\backsection[Data availability statement]{
    The simulation and linking analysis scripts are available at the following repositories:
    \begin{itemize}
        \item Simulation: \url{https://github.com/kburns/active_matter_ball}
        \item Analysis: \url{https://github.com/NicoRomeo/pyfln}
    \end{itemize}
}

\backsection[Author ORCID]{
    N. Romeo, https://orcid.org/0000-0001-6926-5371; 
    J. S\l{}omka https://orcid.org/0000-0002-7097-5810; 
    J. Dunkel https://orcid.org/0000-0001-8865-2369; 
    K. J. Burns, https://orcid.org/0000-0003-4761-4766
}

\appendix
\section{Numerical methods}\label{appA}

We solve the GNS equations Eq.~\eqref{eq:GNS_all}~\&~\eqref{eq:GNS_BCs} using Dedalus v3 \citep{burns_dedalus_2020}.
The variables are discretized using the full-ball basis from \cite{vasil2019tensor,LECOANET2019100012}, consisting of spin-weighted spherical harmonics in the angular directions and one-sided Jacobi polynomials in radius.
This discretization satisfies the exact regularity conditions of smooth tensor fields at the poles and the origin, without having to excise any regions around the coordinate singularities.
Other attempts to simulate the GNS equations in the ball using alternate discretizations have exhibited stability issues \citep{boulle2021optimal}, which may be related to their inexact imposition of regularity conditions at the origin.

In Dedalus, the equations are integrated semi-implicitly with the linear terms solved implicitly and the non-linear computed explicitly using a 3rd-order 4-stage mixed Runge-Kutta (RK) scheme \citep[sec 2.8]{ascher_implicit-explicit_1997}.
The nonlinear terms are computed pseudospectrally with 3/2 dealiasing.
Incompressibility is directly enforced by implicitly solving for the dynamic pressure as a Lagrange multiplier in each RK substep.
The boundary conditions are implemented using a generalized tau method, where radial tau terms are explicitly added to the linear system.
The system is evolved with a fixed timestep from random initial noise until reaching a statistically stationary state.

\begin{figure}
\centerline{\includegraphics{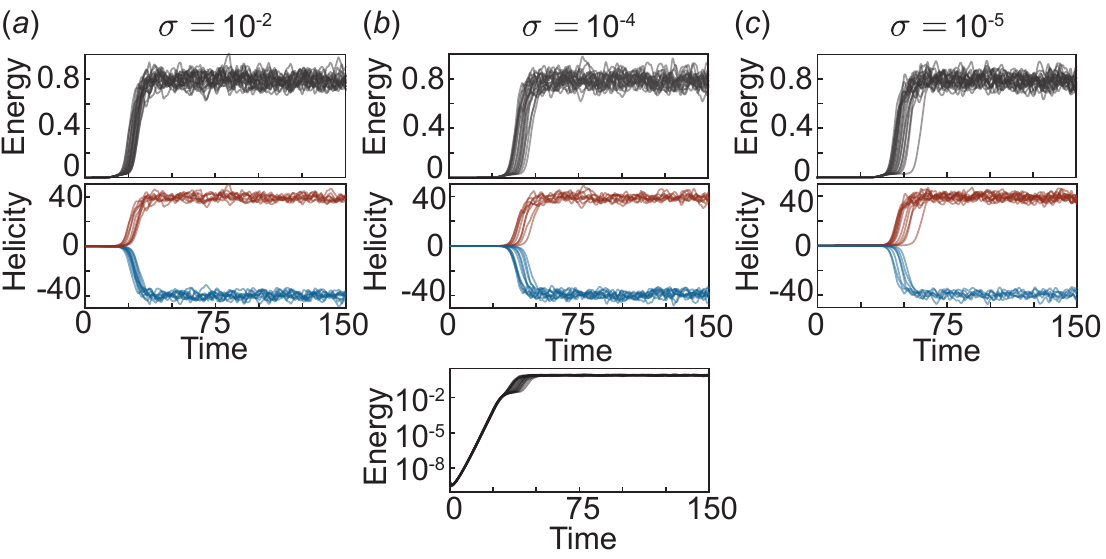}}
\caption{
    \rev{The flow statistics in steady-state are independent of initial condition amplitude: Energy and total helicity as a function of time for random initial conditions with different amplitudes $\sigma$. 
    All simulations have $\Lambda = R/8, \kappa=4/R, \tau = 1$.
    \textit{(a)} 20 simulations with $\sigma = 10^{-2}$.
    \textit{(b)} 20 simulations with $\sigma = 10^{-4}$. The bottom row shows the energy on a semi-logarithmic scale, indicating an exponential growth of the energy at early times, characteristic of a linear instability. 
    \textit{(c)} 20 simulations with $\sigma = 10^{-5}$.
    We note that the main text  results are all reported for $\sigma = 10^{-3}$. 
    All values are in simulation units where $R= 1, \tau = 1$.}
}
\label{fig:fig_amplitude}
\end{figure}

\rev{The initial conditions for velocity are Gaussian white fields with mean value $\langle \bm{u}\rangle = 0$ and spatial correlations between components of the velocity $\langle u_i(\bm{x}) u_j(\bm{y})\rangle = \sigma^2 \delta_{ij}\delta(\bm{x}-\bm{y})$ where $i,j$ index the velocity components.
The random field is then projected onto its incompressible component such that $\nabla \cdot \bm{u} = 0$.
We find that the statistical steady state is independent of the amplitude of the initial amplitude $\sigma$ (figure~\ref{fig:fig_amplitude}), consistent with the fact that the initial transient regime is dominated by a linear instability.}

Vortex line integration is performed using scipy's ODE integrator called through the method $\texttt{solve\_ivp}$. 
The RK45 solver is used with absolute tolerance $10^{-6}$, relative tolerance $10^{-12}$ and maximal step size $10^{-2}$ (in length units where the ball radius is $R=1$).
Linear interpolation of the vorticity field between points on the numerical quadrature grid is used to integrate the vortex lines.

\bibliographystyle{jfm}
\bibliography{jfm_linking}

\end{document}